\documentclass[aps,pra,showpacs,floatfix,amsmath,amssymb,superscriptaddress]{revtex4}
\usepackage{graphicx}
\usepackage{subfigure}

\begin{document}
\title{The squeezed generalized amplitude damping channel}
\author{R. Srikanth}
\email{srik@rri.res.in}
\affiliation{Poornaprajna Institute of Scientific Research,
Devanahalli, Bangalore- 562 110, India}
\affiliation{Raman Research Institute, Sadashiva Nagar,
Bangalore - 560 080, India} 
\author{Subhashish Banerjee}
\email{subhashishb@rri.res.in}
\affiliation{Raman Research Institute, Sadashiva Nagar, Karnataka,
Bangalore - 560 080, India} 

\begin{abstract}
Squeezing of a thermal bath  introduces new features absent in an open
quantum  system  interacting  with  an uncorrelated  (zero  squeezing)
thermal  bath.  The  resulting dynamics,  governed by  a Lindblad-type
evolution,  extends the  concept  of a  generalized amplitude  damping
channel, which corresponds to  a dissipative interaction with a purely
thermal bath.  Here  we present the Kraus representation  of this map,
which we call the  squeezed generalized amplitude damping channel.  As
an application  of this channel  to quantum information, we  study the
classical capacity of this channel.
\end{abstract} 
\pacs{03.65.Yz, 03.67.Hk, 03.67.-a}  

\maketitle

\section{Introduction\label{sec:intro}}

The concept  of open quantum systems  is a ubiquitous one  in that any
real system of interest is surrounded by its environment (reservoir or
bath), which  influences its dynamics.   They provide a  natural route
for discussing damping and dephasing. One of the first testing grounds
for  open  system  ideas   was  in  quantum  optics  \cite{wl73}.  Its
application to other areas gained  momentum from the works of Caldeira
and   Leggett  \cite{cl83},  and   Zurek  \cite{wz93}   among  others.
Depending upon the  system-reservoir ($S-R$) interaction, open systems
can  broadly  be   classified  into  two   categories,  viz.,  quantum
non-demolition  (QND) or  dissipative.  A  particular type  of quantum
nondemolition  (QND)  $S-R$  interaction   is  given  by  a  class  of
energy-preserving  measurements  in  which  dephasing  occurs  without
damping the system.   This may be achieved when  the Hamiltonian $H_S$
of the  system commutes with  the Hamiltonian $H_{SR}$  describing the
system-reservoir  interaction, i.e.,  $H_{SR}$  is a  constant of  the
motion generated  by $H_S$  \cite{sgc96, mp98, gkd01}.   A dissipative
open system would be when  $H_S$ and $H_{SR}$ do not commute resulting
in  dephasing   along  with  damping  \cite{bp02}.    A  prototype  of
dissipative  open quantum  systems, having  many applications,  is the
quantum  Brownian  motion of  harmonic  oscillators.   This model  was
studied by  Caldeira and  Leggett \cite{cl83} for  the case  where the
system  and  its  environment  were initially  separable.   The  above
treatment  of  the quantum  Brownian  motion  was  generalized to  the
physically reasonable initial condition of a mixed state of the system
and  its environment by  Hakim and  Ambegaokar \cite{ha85},  Smith and
Caldeira  \cite{sc87}, Grabert, Schramm  and Ingold  \cite{gsi88}, and
for the case of a system in a Stern-Gerlach potential \cite{sb00}, and
also for the quantum Brownian motion with nonlinear system-environment
couplings \cite{sb03-2}, among others.   The interest in the relevance
of open  system ideas to  quantum information and  quantum computation
has burgeoned in recent times  because of the impressive progress made
on the  experimental front  in the manipulation  of quantum  states of
matter   towards   quantum    information   processing   and   quantum
communication.

A number of open system effects  can be given an operator-sum or Kraus
representation \cite{kraus}.  In  this representation, a superoperator
${\cal  E}$ due  to interaction  with the  environment, acting  on the
state of the system is given by
\begin{equation}
\label{eq:kraus}
\rho  \longrightarrow  {\cal  E}(\rho)  =  \sum_k  \langle  e_k|U(\rho
\otimes |f_0\rangle\langle f_0|)U^{\dag}|e_k\rangle  = \sum_j E_j \rho
E_j^{\dag},
\end{equation}
where $U$ is  the unitary operator representing the  free evolution of
the system,  reservoir, as  well as the  interaction between  the two,
$\{|f_0\rangle\}$   is   the    environment's   initial   state,   and
$\{|e_k\rangle\}$ is a basis  for the environment. The environment and
the system are assumed to start  in a separable state. The $E_j \equiv
\langle e_k|U|f_0\rangle$  are the Kraus operators,  which satisfy the
completeness condition $\sum_j E_j^{\dag}E_j = \mathcal{I}$. It can be
shown  that  any   transformation  that  can  be  cast   in  the  form
(\ref{eq:kraus}) is a completely positive (CP) map \cite{nc00}.

To  connect the predicted  effects to  actual experiments,  a detailed
model  of  the  interaction  between  the  principal  system  and  the
environment is  required. However, from  the viewpoint of a  number of
applications to quantum  computation and information processing, these
details may not be of immediate  relevance.  In such a case, the Kraus
representation is useful because  it provides an intrinsic description
of  the principal system,  without explicitly  considering the
detailed  properties of  the environment  \cite{nc00}.   The essential
features of  the problem are  contained in the operators  $E_k$.  This
not  only  simplifies  calculations,  but often  provides  theoretical
insight. An example  we will encounter below is  the interplay between
environmental  squeezing   and  thermal   effects  for  the   case  of
dissipative  system-reservoir  interactions.   Moreover,  the  reduced
dynamics of  a number of, seemingly different,  physical systems could
be  modelled in the  quantum operations  formalism \cite{nc00}  by the
same noisy channel. This would help in the development of insight into
the common features of the reduced dynamics of the above systems.  For
example, for the case of a two-level system interacting, via a quantum
non-demolition  (QND) interaction,  either  with a  bath of  two-level
systems (in the weak coupling  limit) or harmonic oscillators (at zero
temperature $T$ and zero bath  squeezing), the reduced dynamics in the
quatum operations formalism  can be shown to be  governed by the phase
damping  channel \cite{srigp,  bg06}.   Another example  would be  the
reduced dynamics of a simplified Jaynes-Cummings model consisting of a
two-level  atom coupled  to  a single  cavity  mode which  in turn  is
interacting  with a vacuum  bath of  harmonic oscillators.   This can,
considering  only a single  excitation in  the atom-cavity  system, be
shown to be generated by an amplitude damping channel.  Since as shown
below,  an  amplitude  damping  channel  results  generally  from  the
interactions governed by  the Lindblad type of evolution  (at zero $T$
and zero bath  squeezing), this enables us to  get an understanding of
the reduced  dynamics of  the above system  without having  to concern
ourselves with particular details.

In this paper we study an open system, taken to be two-level system or
qubit, where the  bath is taken to be initially  in a squeezed thermal
state.  The resulting dynamics, governed by a Lindblad-type evolution,
generates  a completely  positive map  that extends  the concept  of a
generalized  amplitude   damping  channel,  which   corresponds  to  a
dissipative interaction  with a purely  thermal bath.  We  present the
Kraus  representation  of  this   map,  which  we  call  the  squeezed
generalized  amplitude  damping  channel.   An advantage  of  using  a
squeezed thermal bath is that  the decay rate of quantum coherence can
be  suppressed   leading  to  preservation   of  nonclassical  effects
\cite{bg06,kw88,kb93}. It has also  been shown to modify the evolution
of the geometric phase  of two-level atomic systems \cite{srigp}.  The
preservation of  entanglement in the  presence of a squeezed  bath has
been investigated in Ref. \cite{wlk02}.

The paper  is organized as  follows.  In Section  \ref{sec:nonqnd}, we
obtain the evolution of a qubit in a dissipative (non-QND) interaction
with its bath.  In  Section \ref{sec:gpbma}, we consider, in specific,
a  system  interacting  with  a  squeezed thermal  bath  in  the  weak
Born-Markov rotating wave  approximation.  In Section \ref{sec:jc}, we
consider a  single-mode Jaynes-Cummings model  in a vacuum  bath.  The
amplitude  damping  and  generalized  amplitude damping  channels  are
introduced in Section \ref{sec:gad}, where  it is pointed out that the
simplified  Jaynes-Cummings   model  realizes  an   amplitude  damping
channel, while the weak Born-Markov interaction without bath squeezing
realizes a  generalized amplitude  damping channel.  We  introduce the
squeezed  generalized  amplitude damping  channel,  which extends  the
concept  of generalized  amplitude damping  noise, to  the  case where
environmental  squeezing is  included, in  Section  \ref{sec:bma}.  Of
particular  interest is  the fact  that unlike  the case  of  a purely
dephasing channel,  where the action of squeezing  and temperature are
concurrently decohering, in the case of squeezed generalized amplitude
damping   channel,    they   can   exhibit    counteractive   behavior
\cite{srigp,sriph}.   In specific, in  Section \ref{sec:cc},  where we
study  the  classical capacity  of  a  squeezed generalized  amplitude
damping  channel,  we show  that  squeezing  can  improve the  channel
capacity, whereas  temperature necessarily  degrades it.  We  make our
conclusions in Section \ref{sec:konklu}.

\section{Two-level system in non-QND interaction with bath
\label{sec:nonqnd}}

In  this section  we study  the dynamics  of a  two-level system  in a
dissipative interaction with its bath,  which is taken as one composed
of harmonic  oscillators.  We  first consider the  case of  the system
interacting  with a  bath which  is  initially in  a squeezed  thermal
state, in the weak  coupling Born-Markov, rotating wave approximation.
Next  we consider  a simple  single  mode Jaynes-Cummings  model in  a
vacuum bath.

\subsection{System interacting with bath in the weak Born-Markov,
rotating-wave approximation \label{sec:gpbma}}

Here we  take up  the case  of a two-level  system interacting  with a
squeezed  thermal   bath  in  the  weak   Born-Markov,  rotating  wave
approximation.    The  system   Hamiltonian   is  given   by  $H_S   =
(\hbar\omega/2)\sigma_z$.   The  system  interacts  with the  bath  of
harmonic  oscillators via  the  atomic dipole  operator  which in  the
interaction  picture  is  given  as  $\vec{D}(t)  =  \vec{d}  \sigma_-
e^{-i\omega t} + \vec{d^*}  \sigma_+ e^{i\omega t}$ where $\vec{d}$ is
the transition matrix elements  of the dipole operator.  The evolution
of  the reduced  density  matrix operator  of  the system  $S$ in  the
interaction picture has the following form \cite{sz97, bp02}
\begin{eqnarray}
{d \over dt}\rho^s(t) &=& \gamma_0 (N + 1) \left(\sigma_-  \rho^s(t)
\sigma_+ - {1 \over 2}\sigma_+ \sigma_- \rho^s(t) -
{1 \over 2} \rho^s(t) \sigma_+ \sigma_- \right) \nonumber\\
& + & \gamma_0 N \left( \sigma_+  \rho^s(t)
\sigma_- - {1 \over 2}\sigma_- \sigma_+ \rho^s(t) -
{1 \over 2} \rho^s(t) \sigma_- \sigma_+ \right) \nonumber\\
& - & \gamma_0 M   \sigma_+  \rho^s(t) \sigma_+ -
\gamma_0 M^* \sigma_-  \rho^s(t) \sigma_-. 
\label{4b} 
\end{eqnarray}
Here $\gamma_0$ is the spontaneous  emission rate given by $\gamma_0 =
(4  \omega^3 |\vec{d}|^2)/(3 \hbar  c^3)$, and  $\sigma_+$, $\sigma_-$
are the standard raising and lowering operators, respectively given by
$\sigma_+ =  |1 \rangle  \langle 0| =  \frac{1}{2} \left(\sigma_x  + i
\sigma_y \right)$ and $\sigma_- = |0  \rangle \langle 1| = {1 \over 2}
\left(\sigma_x - i \sigma_y \right)$.  Eq. (\ref{4b}) may be expressed
in a manifestly Lindblad form as \cite{srigp}
\begin{equation}
\frac{d}{dt}\rho^s(t) = \sum_{j=1}^2\left(
2R_j\rho^s R^{\dag}_j - R_j^{\dag}R_j\rho^s - \rho^s R_j^{\dag}R_j\right),
\label{eq:lindblad}
\end{equation}
where    $R_1   =    (\gamma_0(N_{\rm   th}+1)/2)^{1/2}R$,    $R_2   =
(\gamma_0N_{\rm  th}/2)^{1/2}R^{\dag}$  and  $R =  \sigma_-\cosh(r)  +
e^{i\Phi}\sigma_+\sinh(r)$.  This   observation  guarantees  that  the
evolution of the density operator can be given a Kraus or operator-sum
representation  \cite{nc00},   a  point   we  return  to   in  Section
\ref{sec:bma}. If  $T=0$, then $R_2$  vanishes, and a  single Lindblad
operator suffices to describe Eq. (\ref{4b}).

A number of methods of generating bath squeezing have been proposed in
the literature.  A squeezed reservoir  may be constructed on the basis
of establishment of squeezed  light field \cite{buz91}.  A single-mode
squeezed state  created in a degenerate  parametric amplifier operated
in  an appropriate  cavity,  when  coupled to  an  infinite number  of
external ``output''  modes, transfers the  squeezing into correlations
between  side-bands  of the  multimode  light  field.  A  subthreshold
optical parametric oscillator  (OPO) can be used as  the basis for the
implementation  of  a   stable,  reliable  source  of  continuous-wave
squeezed    vacuum     \cite{bre95}.     Experiments    probing    the
squeezed-light-atom   system   have   been   carried  out   in   Refs.
\cite{geo95,tur98}.   In particular,  the latter  reference  details a
method in  which an OPO  operated below threshold downconverts  a high
energy  photon into  two correlated  low energy  photons  generating a
close-to-minimum-uncertainty squeezed  vacuum state. At  the output of
the OPO, the  squeezed vacuum is mixed on a  99-1 beam-splitter with a
phase-coherent reference oscillator  with controlled relative phase to
the squeezing,  resulting in a combined electromagnetic  field that is
equivalent to a displaced squeezed state.

An infinite  array of beam-splitters can  be used to  model a squeezed
reservoir \cite{kim95}.  The  signal is injected from the  left of the
array, with independent squeezed fields (all with the same properties)
injected into  the other  ports. The output  of a  given beam-splitter
serves as  input to the subsequent  one.  In the limit  of an infinite
number of beam-splitters, the dynamics generated by Eq.  (\ref{4b}) is
simulated.  In Ref.   \cite{poy96}, quantum reservoir engineering with
laser-cooled trapped  ions is  used to mimic  the dynamics of  an atom
interacting  with  a  squeezed   vacuum  bath  (Eq.   (\ref{4b})  with
temperature  $T$ set  to  zero).  Another way  to  engineer a  quantum
reservoir to mimic coupling with  a squeezed bath has been proposed in
Ref.   \cite{lut98}.   They  consider  a four-level  system  with  two
(degenerate) ground  and excited states,  driven by weak  laser fields
and coupled  to a vacuum  reservoir of radiation  modes.  Interference
between the spontaneous emission  channels in optical pumping leads to
a squeezed bath type coupling  for the two-level system constituted by
the two ground levels.  The  properties of the squeezed bath are shown
to be controllable by means  of the laser parameters.  An experimental
study  of the decoherence  and decay  of quantum  states of  a trapped
atomic  ion's  harmonic  motion   with  several  types  of  engineered
reservoirs,  including amplitude  dampling  and generalized  amplitude
damping channels, have been made in Refs. \cite{mya00,tur00}.

In Eqs.  (\ref{4b}) and  (\ref{eq:lindblad}), we use  the nomenclature
$|1 \rangle$ for the upper state  and $|0 \rangle$ for the lower state
and $\sigma_x,  \sigma_y, \sigma_z$  are the standard  Pauli matrices.
In Eq.  (\ref{4b}),
\begin{equation}
\label{eq:N}
N   =  N_{\rm   th}(\cosh^2(r)  +   \sinh^2(r))  +
\sinh^2(r),
\end{equation}
and $M = -\frac{1}{2} \sinh(2r) e^{i\Phi} (2 N_{\rm th} +
1)$, and $N_{\rm th} =  1/(e^{\hbar \omega/k_B T} - 1)$.  Here $N_{\rm
th}$ is the  Planck distribution giving the number  of thermal photons
at the  frequency $\omega$ and  $r$, $\Phi$ are  bath squeezing parameters
\cite{cs85}.
The analogous case of a thermal bath without squeezing can be obtained
from the  above expressions by  setting these squeezing  parameters to
zero.  We solve the Eq. (\ref{4b}) using the Bloch vector formalism as
\begin{equation}
\rho^s (t) = {1 \over 2} \left(\mathcal{I} + \langle 
\vec{\sigma}(t)\rangle\cdot\vec{\sigma}
\right)
= \begin{pmatrix}
{1 \over 2} \left(1 + \langle \sigma_z(t) \rangle
\right) & \langle \sigma_-(t) \rangle \cr
\langle \sigma_+(t) \rangle & {1 \over 2} \left(1 -
\langle \sigma_z(t) \rangle \right)
\end{pmatrix}. \label{4g} 
\end{equation}
In   Eq.  (\ref{4g})   by   the  vector   $\vec{\sigma}(t)$  we   mean
$(\sigma_x(t), \sigma_y(t), \sigma_z(t))$ and $\langle \vec{\sigma}(t)
\rangle$   denotes  the   Bloch   vectors  which   are  solved   using
Eq. (\ref{4b}) to yield \cite{srigp}
\begin{eqnarray}
\label{eq:bmbloch}
\langle \sigma_x (t) \rangle &=& \left[1 + {1 \over 2} \left(e^{\gamma_0 a t}
- 1\right) (1 + \cos(\Phi))\right] e^{-{\gamma_0 \over 2}(2N + 1 + a)t}
\langle \sigma_x (0) \rangle 
- \sin(\Phi) \sinh({\gamma_0 a t \over 2}) e^{-{\gamma_0 \over 2}(2N + 1)t}
\langle \sigma_y (0) \rangle, \nonumber \\
\langle \sigma_y (t) \rangle &=& \left[1 + {1 \over 2} \left(e^{\gamma_0 a t}
- 1\right) (1 - \cos(\Phi))\right] e^{-{\gamma_0 \over 2}(2N + 1 + a)t}
\langle \sigma_y (0) \rangle 
- \sin(\Phi) \sinh({\gamma_0 a t \over 2}) e^{-{\gamma_0 \over 2}(2N + 1)t}
\langle \sigma_x (0) \rangle, \nonumber \\
\langle \sigma_z (t) \rangle &=& e^{-\gamma_0 (2N + 1)t} \langle 
\sigma_z (0) \rangle - {1 \over (2N + 1)} \left(1 - e^{-\gamma_0 (2N + 1)t} 
\right). 
\end{eqnarray}
In these equations $a = \sinh(2r) (2N_{th} + 1)$.
Using the Eqs. (\ref{eq:bmbloch}) in Eq. (\ref{4g}) 
and then reverting back to the Schr\"{o}dinger picture, the
reduced density matrix of the system can be written as
\begin{equation}
\label{eq:bmrhos}
\rho^s (t) = \begin{pmatrix} {1 \over 2} (1 + A) & B e^{-i \omega t} 
\cr B^* e^{i \omega t} & 
{1 \over 2} (1 - A)
\end{pmatrix}, 
\end{equation}
where, in view of Eq. (\ref{4g}),
\begin{equation}
A \equiv \langle\sigma_z(t)\rangle
= e^{-\gamma_0 (2N + 1)t} \langle 
\sigma_z (0) \rangle - {1 \over (2N + 1)} \left(1 - e^{-\gamma_0 (2N + 1)t} 
\right), \label{4m} 
\end{equation}
\begin{equation}
B = \left[1 + {1 \over 2} \left(e^{\gamma_0 a t}
- 1\right) \right] e^{-{\gamma_0 \over 2}(2N + 1 + a)t}
\langle \sigma_- (0) \rangle
+ \sinh({\gamma_0 a t \over 2}) e^{i \Phi 
- {\gamma_0 \over 2}(2N + 1)t} \langle \sigma_+ (0) \rangle. \label{4n}
\end{equation}

From Eq. (\ref{eq:bmbloch}), it is seen that the system evolves
towards a fixed asymptotic point in the Bloch sphere \cite{bg06}, 
which in general is not a pure state,
but the mixture
\begin{equation}
\rho_{\rm asymp}
= \left(\begin{array}{ll} 1-q & 0 \\ 0 & q \end{array} \right),
\end{equation}
where $q = \frac{N+1}{2N+1}$.
If $T=0$ and $r=0$, then $q=1$, and the pure state $|0\rangle$
is reached asymptotically, an observation that serves as the basis
for the quantum deleter \cite{qdele}. 

\subsection{Simplified Jaynes-Cummings Model\label{sec:jc}}

Here  we  consider  a  simplified Jaynes-Cummings  model  taking  into
account the  effect of  the environment, which  is modelled as  a zero
temperature bath.  In this model we consider the case of only a single
excitation  in the  atom-cavity  system with  the  bath modelling  the
effect  of imperfect  cavity  mirrors. Also  the  cavity frequency  is
assumed  to be  in  resonance with  the  atomic frequency  \cite{bp02,
bg97}. The total Hamiltonian is
\begin{equation}
H  =  H_S + H_R + H_{SR} = \omega_0 \sigma_+ \sigma_- + 
\sum\limits_k \omega_k b^{\dagger}_k b_k +
\sigma_+ \sum\limits_k g_k b_k + \sigma_- \sum\limits_k g^*_k b^{\dagger}_k.
\label{4b1} 
\end{equation} 
Here  $H_S$, $H_R$  and $H_{SR}$  stand  for the  Hamiltonians of  the
system, reservoir and  system reservoir interaction, respectively.  In
the case of a single  excitation in the atom-cavity system, the cavity
mode can be  eliminated in favour of the  following effective spectral
density
\begin{equation}
I(\omega) = {1 \over 2\pi} {\gamma_0 \kappa^2 \over (\omega_0 - \omega)^2
+ \kappa^2}. \label{4b2}
\end{equation}
Here $\omega_0$ is the atomic transition frequency and $\kappa$ is the
spectral width  of the  system-environment coupling. Tracing  over the
vacuum  bath and  assuming that  initially there  are no  photons, the
reduced density matrix of the  atom (two-level system) can be obtained
in the Schr{\"o}dinger representation as
\begin{equation}
\rho^s (t) = \begin{pmatrix} a  & b e^{- i \omega_0 t} \cr 
b^*  e^{i \omega_0 t} & 
(1 - a)
\end{pmatrix}, \label{4b3}
\end{equation}
where
\begin{subequations}
\label{eq:jc}
\begin{eqnarray}
a &=& \rho_{{1 \over 2}, {1 \over 2}}(0) 
e^{-\kappa t}\left[\cosh\left({l t \over 2}\right)
+ {\kappa \over l} \sinh\left({l t \over 2}\right) \right]^2, \label{4b4} 
\\
b &=& \rho_{{1 \over 2}, -{1 \over 2}}(0) e^{-{\kappa t \over 2}}
\left[\cosh\left({l t \over 2}\right)
+ {\kappa \over l} \sinh\left({l t \over 2}\right) \right]. \label{4b5} 
\end{eqnarray}  
\end{subequations}
Here $l = \sqrt{\kappa^2 - 2 \gamma_0 \kappa}$, where $\gamma_0, \kappa$
are as in Eq. (\ref{4b2}). Initially the system is chosen to be in the state
\begin{equation}
|\psi(0)\rangle = \cos({\theta_0 \over 2}) |1\rangle + e^{ i \phi_0}
\sin({\theta_0 \over 2}) |0\rangle. 
\label{4b6}
\end{equation}
From the above equation, it can be easily seen that 
$\rho_{{1 \over 2},{1 \over 2}} (0) = \cos^2({\theta_0 \over 2})$ and
$\rho_{{1 \over 2},-{1 \over 2}}(0) = {1 \over 2}  e^{-i \phi_0}
\sin(\theta_0)$. 

\section{Amplitude damping and 
Generalized Amplitude damping channels \label{sec:gad}}

The generalized amplitude channel  is generated by the evolution given
by the master equation  (\ref{4b}), with the bath squeezing parameters
$r$ and  $\Phi$ set to  zero.  Generalized amplitude  damping channels
capture the idea of energy  dissipation from a system, for example, in
the spontaneous  emission of a photon,  or when a spin  system at high
temperature   approaches  equilibrium   with  its   environment  (also
cf. Refs. \cite{mya00,tur00}). A  simple model of an amplitude damping
channel is the scattering of a  photon via a beam-splitter. One of the
output modes  is the  environment, which is  traced out.   The unitary
transformation at the beam-splitter  is given by $B = \exp\left[\theta
(a^{\dag}b - ab^{\dag})\right]$,  where $a, b$ and $a^{\dag},b^{\dag}$
are the  annihilation and  creation operators for  photons in  the two
modes.

\subsection{Amplitude damping channel \label{sec:ampdamp}}

This  channel  is generated  by  the  evolution  given by  the  master
equation   (\ref{4b}),  with  temperature   $T$  and   bath  squeezing
parameters  $r$  and  $\Phi$  set  to zero.  The  corresponding  Kraus
operators are:
\begin{eqnarray}
\label{eq:gbmakraus1}
\begin{array}{ll}
E_0 \equiv \left[\begin{array}{ll} 
\sqrt{1-\lambda(t)} & 0 \\ 0 & 1
\end{array}\right]; ~~~~ &
E_1 \equiv \left[\begin{array}{ll} 0 & 0 \\ \sqrt{\lambda(t)} & 0
\end{array}\right].
\end{array}
\end{eqnarray}

The effect of these operators is to produce the completely positive map
\begin{equation}
\sum_j E_j \left( \begin{array}{ll} A & B^* \\ B & 1-A
\end{array}\right) E^{\dag}_j
= \left( \begin{array}{ll} A(1-\lambda) & \sqrt{1-\lambda}B^* 
\\ \sqrt{1-\lambda}B & 1-A + \lambda A 
\end{array}\right),
\label{eq:map1}
\end{equation}
where,  on  comparison  with  Eq.   (\ref{4b6}),  we  see  that  $A  =
\cos^2(\theta_0/2)$,   $B   =  (1/2)e^{i\phi_0}\sin(\theta_0)$.    The
simplified Jaynes-Cummings model of  the previous subsection is easily
seen to  realize an amplitude damping channel.   It is straightforward
to verify that with the identification
\begin{equation}
\label{eq:lambda0}
1-\lambda(t) \equiv e^{-{\kappa t}}
\left[\cosh\left({l t \over 2}\right)
+ {\kappa \over l} \sinh\left({l t \over 2}\right) \right]^2. 
\end{equation}
the operators (\ref{eq:gbmakraus1}),  acting on the state (\ref{4b6}),
reproduce the evolution (\ref{4b3}) (in the interaction picture).

\subsection{Generalized amplitude damping channel \label{sec:gnampdamp}}

This  channel  is generated  by  the  evolution  governed
by  the  master
equation  (\ref{4b}), with  bath squeezing  parameters $r$ 
and $\Phi$  set to
zero, but $T$ not necessarily zero.   The corresponding Kraus
operators are:
\begin{eqnarray}
\label{eq:gbmakraus}
\begin{array}{ll}
E_0 \equiv \sqrt{p}\left[\begin{array}{ll} 
\sqrt{1-\lambda(t)} & 0 \\ 0 & 1
\end{array}\right]; ~~~~ &
E_1 \equiv \sqrt{p}\left[\begin{array}{ll} 0 & 0 \\ \sqrt{\lambda(t)} & 0
\end{array}\right];  \\
E_2 \equiv \sqrt{1-p}\left[\begin{array}{ll} 1 & 0 \\ 0 & 
\sqrt{1-\lambda(t)}
\end{array}\right]; ~~~~ &
E_3 \equiv \sqrt{1-p}\left[\begin{array}{ll} 0 & \sqrt{\lambda(t)} \\ 0 & 0
\end{array}\right],
\end{array}
\end{eqnarray}
where $0 \le p \le 1$ \cite{nc00,srigp}. 

The effect of these operators is to produce the completely positive map
\begin{eqnarray}
\sum_j E_j \left( \begin{array}{ll} A & B^* \\ B & 1-A
\end{array}\right) E^{\dag}_j
&=& p \left( \begin{array}{ll} A(1-\lambda) & \sqrt{1-\lambda}B^* 
\\ \sqrt{1-\lambda}B & 1-A + \lambda A 
\end{array}\right) \nonumber \\
& & + (1-p) \left( \begin{array}{ll} A + \lambda(1-A) & \sqrt{1-\lambda}B^* 
\\ \sqrt{1-\lambda}B & 1-A + (1-\lambda)(1-A)
\end{array}\right).
\label{eq:map}
\end{eqnarray}
It is straightforward to verify that with the identification
\begin{equation}
\label{eq:lambda}
\lambda(t) \equiv 1 - e^{-\gamma_0(2N_{\rm th} +1) t};\hspace{1.0cm} p
\equiv \frac{N_{\rm th}+1}{2N_{\rm th} +1},
\end{equation}
the  operators  (\ref{eq:gbmakraus})   acting  on   the  state
(\ref{4b6})    reproduce   the   evolution    (\ref{eq:bmbloch}), 
with squeezing set  to zero but
temperature non-vanishing,  by means of the  map 
given by Eq. (\ref{eq:kraus}).
If   $T=0$,  then   $p=1$,  reducing   Eq.    (\ref{eq:gbmakraus})  to
the amplitude damping channel, given by Eq. (\ref{eq:gbmakraus1}).

\section{The squeezed generalized amplitude damping channel \label{sec:bma}}

This  channel  is generated  by  the  evolution  given by  the  master
equation (\ref{4b}), with neither the bath squeezing parameters $r$
and $\Phi$
nor the temperature $T$ necessarily  zero. Thus this is a very general
(completely positive) map generated by Eq.  (\ref{4b}).  To generalize
(\ref{eq:gbmakraus}) to include the effects of squeezing, we construct
the following set of Kraus operators:
\begin{eqnarray}
\begin{array}{ll}
E_0 \equiv \sqrt{p_1}\left[\begin{array}{ll} 
\sqrt{1-\alpha(t)} & 0 \\ 0 & \sqrt{1-\beta(t)}
\end{array}\right]; ~~~~ &
E_1 \equiv \sqrt{p_1}\left[\begin{array}{ll} 0 & \sqrt{\beta(t)} 
\\ \sqrt{\alpha(t)}e^{-i\phi(t)}  & 0 \end{array}\right];  \\
E_2 \equiv \sqrt{p_2}\left[\begin{array}{ll} 
\sqrt{1-\mu(t)} & 0 \\ 0 & \sqrt{1-\nu(t)}
\end{array}\right]; ~~~~ &
E_3 \equiv \sqrt{p_2}\left[\begin{array}{ll} 0 & \sqrt{\nu(t)}
\\ \sqrt{\mu(t)}e^{-i\theta(t)}  & 0
\end{array}\right].  \\
\end{array}
\label{eq:new}
\end{eqnarray}
It is readily checked that Eq. (\ref{eq:new}) satisfies the
completeness condition
\begin{equation}
\sum_{j=0}^3 E_j^{\dag}E_j = \mathbb{I},
\end{equation}
provided
\begin{equation}
\label{eq:p1p2}
p_1 + p_2= 1.
\end{equation}

Substituting the  Kraus operator elements given  by Eq. (\ref{eq:new})
in Eq.   (\ref{eq:kraus}), and using Eq. (\ref{4g}),
yields the following  Bloch vector evolution equations:
\begin{subequations}
\label{eq:newev}
\begin{eqnarray}
\langle\sigma_x(t)\rangle &=&
\left[(p_1\sqrt{(1-\alpha(t))(1-\beta(t))})+ p_2\sqrt{(1-\mu(t))(1-\nu(t))} 
+ (p_1\sqrt{\alpha(t)\beta(t)}\cos\phi + p_2\sqrt{\mu(t) \nu(t)}\cos\theta )\right]
\langle\sigma_x(0)\rangle \nonumber \\
&-& \left[( p_1\sqrt{\alpha(t)\beta(t)}\sin\phi +
   p_2\sqrt{\mu(t) \nu(t)}\sin\theta 
   )\right]\langle\sigma_y(0)\rangle, 
   \label{eq:newev1} \\ 
\langle\sigma_y(t)\rangle &=&
\left[(p_1\sqrt{(1-\alpha(t))(1-\beta(t))} + p_2\sqrt{(1-\mu(t))(1-\nu(t))} )
- (p_1\sqrt{\alpha(t)\beta(t)}\cos\phi + p_2\sqrt{\mu(t) \nu(t)}\cos\theta 
   )\right]\langle\sigma_y(0)\rangle \nonumber\\
&-& \left[( p_1\sqrt{\alpha(t)\beta(t)}\sin\phi + p_2\sqrt{\mu(t) \nu(t)}\sin\theta 
   )\right]\langle\sigma_x(0)\rangle, 
   \label{eq:newev2} \\ 
\langle\sigma_z(t)\rangle &=& 
   (1 - p_2(\mu(t)+\nu(t)) - p_1(\alpha(t)+\beta(t)))\langle\sigma_z(0)\rangle 
  - p_2(\mu(t)-\nu(t)) - p_1(\alpha(t)-\beta(t)).
\end{eqnarray}
\end{subequations}

Comparing Eqs.  (\ref{eq:newev}) with Eqs.  (\ref{eq:bmbloch}), we can
read   off  the  corresponding   terms.   In   fact,  the   system  is
underdetermined  as there  are  more variables  than constraints.   An
inspection of Eqs. (\ref{eq:bmbloch}) shows that they yield a total of
5 constraints  on the channel  variables, $p_1$, $p_2$,  $\mu$, $\nu$,
$\alpha$,  $\beta$, $\theta$  and  $\phi$, with  a further  constraint
coming from Eq.  (\ref{eq:p1p2}).   The two redundant variables may be
conveniently chosen to be $\beta$ and $\phi$.  Setting $\beta=\phi=0$,
a comparison of  Eqs. (\ref{eq:newev}) and (\ref{eq:bmbloch}) produces
the following relations:
\begin{subequations}
\label{eq:assoc}
\begin{eqnarray}
p_1\sqrt{(1-\alpha(t))} + p_2\sqrt{(1-\mu(t))(1-\nu(t))} 
&=& 
\cosh\left(\frac{\gamma_0 at}{2}\right)\exp\left(-\frac{\gamma_0}{2}
(2N+1)t\right),
\label{eq:assoc1} \\
p_2\sqrt{\mu(t)\nu(t)}\cos\theta &=& \cos(\Phi)
\sinh\left(\frac{\gamma_0 at}{2}\right)\exp\left(-\frac{\gamma_0}{2}
(2N+1)t\right),
\label{eq:assoc2} \\
p_2\sqrt{\mu(t)\nu(t)}\sin\theta &=& 
\sin(\Phi) \sinh\left(\frac{\gamma_0 a t}{2}\right)\exp\left(-\frac{\gamma_0}{2}
(2N+1)t\right),
\label{eq:assoc3}\\
p_1\alpha(t) + p_2(\mu(t)-\nu(t)) &=&
\frac{1}{(2N + 1)} 
\left(1 - e^{-\gamma_0 (2N + 1)t}\right),\label{eq:assoc4} \\  
1-
p_1\alpha(t) - p_2(\mu(t)+\nu(t)) 
&=&
e^{-\gamma_0 (2N + 1)t},\label{eq:assoc5} 
\end{eqnarray}
\end{subequations}
and the  required  squeezed  generalized
amplitude damping channel is given, in place of Eq. (\ref{eq:new}), by
the Kraus operators
\begin{eqnarray}
\begin{array}{ll}
E_0 \equiv \sqrt{p_1}\left[\begin{array}{ll} 
\sqrt{1-\alpha(t)} & 0 \\ 0 & 1
\end{array}\right]; ~~~~ &
E_1 \equiv \sqrt{p_1}\left[\begin{array}{ll} 0 & 0 
\\ \sqrt{\alpha(t)} & 0 \end{array}\right];  \\
E_2 \equiv \sqrt{p_2}\left[\begin{array}{ll} 
\sqrt{1-\mu(t)} & 0 \\ 0 & \sqrt{1-\nu(t)}
\end{array}\right]; ~~~~ &
E_3 \equiv \sqrt{p_2}\left[\begin{array}{ll} 0 & \sqrt{\nu(t)} 
\\ \sqrt{\mu(t)}e^{-i\theta(t)}  & 0
\end{array}\right]. 
\end{array}
\label{eq:new0}
\end{eqnarray}

It  is   seen  from  Eqs.    (\ref{eq:assoc})  that  at   time  $t=0$,
$\mu(0)=\nu(0)=\alpha(0)=0$.  We  now determine the  remaining channel
parameters.   From Eqs.   (\ref{eq:assoc2}) and  (\ref{eq:assoc3}), we
find
\begin{equation}
\tan\theta = \tan\Phi, \label{theta}
\end{equation}
allowing  us  to  set  $\theta=\Phi$,  and  to  identify  the  channel
parameter  $\theta$  with  the  bath squeezing  angle.  The  remaining
channel parameters may be identified as follows.

From Eqs. (\ref{eq:assoc4}) and (\ref{eq:assoc5}), we find
\begin{equation}
\nu(t) = \frac{N}{p_2(2N+1)}(1-e^{-\gamma_0(2N+1)t}).
\label{eq:nu}
\end{equation}
Substituting Eq. (\ref{eq:nu}) into Eq. (\ref{eq:assoc3}), we find
\begin{equation}
\mu(t) = \frac{2N+1}{2p_2 N}
\frac{\sinh^2(\gamma_0at/2)}{\sinh(\gamma_0(2N+1)t/2)}
\exp\left(-\frac{\gamma_0}{2}(2N+1)t\right).
\label{eq:mu}
\end{equation}
Using Eqs. (\ref{eq:nu}) and (\ref{eq:mu}) in
(\ref{eq:assoc5}), we obtain
\begin{equation}
\label{eq:tau}
\alpha(t) = \frac{1}{p_1}\left(1 - p_2[\mu(t)+\nu(t)]
- e^{-\gamma_0(2N+1)t}\right),
\end{equation}
where  $\nu(t)$  and $\mu(t)$  are  given  by  Eqs. (\ref{eq:nu})  and
(\ref{eq:mu}), respectively.

Substituting  Eqs.  (\ref{eq:nu}),  (\ref{eq:mu}), (\ref{eq:tau})  and
(\ref{eq:p1p2})  into  Eq.  (\ref{eq:assoc1}),  we obtain  after  some
manipulations
\begin{eqnarray}
p_2(t) &=& \frac{1}{(A+B-C-1)^2-4D}
 \times \left[A^2B + C^2 + A(B^2 - C - B(1+C)-D) - (1+B)D - C(B+D-1) 
\nonumber \right. \\
&\pm& \left.  2\sqrt{D(B-AB+(A-1)C+D)(A-AB+(B-1)C+D)}\right],
\label{eq:p2}
\end{eqnarray}
where
\begin{eqnarray}
A &=& \frac{2N+1}{2N} \frac{\sinh^2(\gamma_0 at/2)}
{\sinh(\gamma_0(2N+1)t/2)}
\exp\left(-\gamma_0(2N+1)t/2\right), \nonumber \\
B &=& \frac{N}{2N+1}(1-\exp(-\gamma_0(2N+1)t)), \nonumber \\
C &= & A + B + \exp(-\gamma_0 (2N+1)t), \nonumber \\
D &=& \cosh^2(\gamma_0 at/2)\exp(-\gamma_0(2N+1)t).
\label{eq:auxip2}
\end{eqnarray}

If the squeezing parameter $r$ is  
set  to  zero,  then $a=0$,  and  it  follows  from
Eq. (\ref{eq:p2}) and (\ref{eq:auxip2}), that
$\hat{p}_2 = N_{\rm th}/(2N_{\rm th}+1)$, where the hat indicates that
squeezing  has   been  set  to  zero.    It  can  be   seen  from  Eq.
(\ref{eq:mu}),  that  for  zero squeezing  ($a=0$),  $\hat{\mu}(t)=0$.
Substituting   $\hat{p}_2  =  N_{\rm   th}/(2N_{\rm  th}+1)$   in  Eq.
(\ref{eq:nu}),   we   find   $\hat{\nu}(t)  =   1-e^{-\gamma_0(2N_{\rm
th}+1)t}$.  Now, a comparison of  Eqs.    (\ref{eq:new0})  with  Eqs.
(\ref{eq:gbmakraus}) shows that $\hat{\nu}(t)=\lambda(t)$,  given by
Eq.   (\ref{eq:lambda}).   

Further, in  view  of Eq.  (\ref{eq:p1p2}),
$\hat{p}_1 =  (N_{\rm th}+1)/(2N_{\rm th}+1)$.
Substituting Eq.   (\ref{eq:nu}) and
the conditions $a=0$ and  $\hat{\mu}(t)=0$ into Eq. (\ref{eq:tau}), it
is easily verified that
\begin{eqnarray}
\hat{\alpha}(t) &=& \frac{1}{p_1}\left(1 - p_2[\hat{\mu}(t)+\hat{\nu}(t)]
- e^{-\gamma_0(2N_{\rm th}+1)t}\right) \nonumber \\
&=& \frac{2N_{\rm th}+1}{N_{\rm th}+1}\left(\hat{\nu}(t) - 
\frac{N_{\rm th}}{2N_{\rm th}+1}\hat{\nu}(t)\right) \nonumber \\
&=& \frac{2N_{\rm th}+1}{N_{\rm th}+1}\left(\frac{N_{\rm th}+1}{2N_{\rm th}+1}
\right)\hat{\nu}(t) \nonumber \\
&=& \hat{\nu}(t).
\label{eq:ha}
\end{eqnarray}
We thus have $\hat{\alpha}(t) = \hat{\nu}(t) = \lambda(t)$, and 
hence the generalized amplitude damping channel 
(\ref{eq:gbmakraus}) is recovered
from Eq. (\ref{eq:new0}) in the limit of vanishing squeezing.

\begin{figure}
\includegraphics[width=8.0cm]{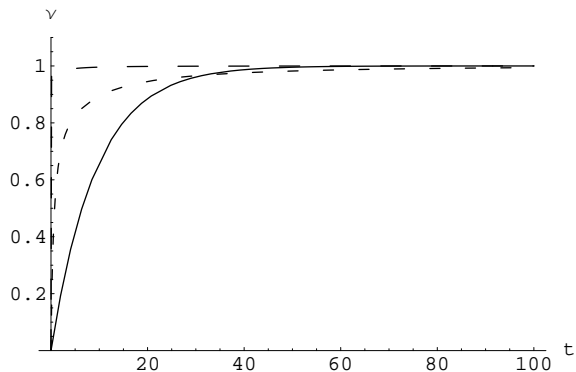}
\caption{$\nu(t)$  [Eq.  (\ref{eq:nu})]  with  respect  to  $t$,  with
$\gamma_0=0.05$. Asymptotically, $\nu(t)$ reaches the value
1. The solid and  small-dashed curves corresponds to the
temperature (in units where $\hbar\equiv k_B\equiv 1$) $T=1$, with the
bath  squeezing   parameter  $r=$   0,  1,  respectively,   while  the
large-dashed curve corresponds to $T=3$ and $r=1$.}
\label{fig:nu}
\end{figure}

\begin{figure}
\includegraphics[width=8.0cm]{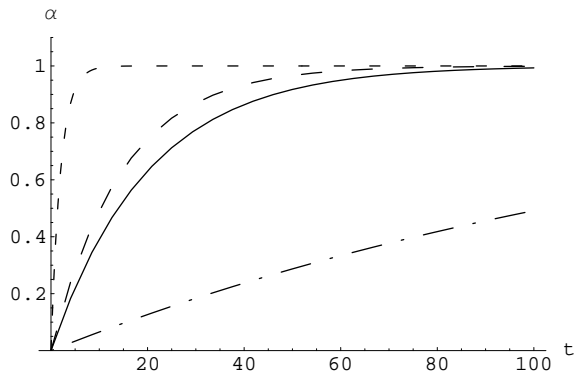}
\caption{$\alpha(t)$  [Eq. (\ref{eq:tau})] with  respect to  time $t$,
with  $\gamma_0=0.05$,  bringing   out  the  counteractive  effect  of
squeezing on  temperature.  Asymptotically, $\alpha(t)$ reaches 1.
We find that  increasing squeezing reduces
$\alpha$  at   any  fixed   temperature,  and  thus   counteracts  the
thermal  effects.  The solid
and  dot-dashed  curves  correspond  to temperature  (in  units  where
$\hbar\equiv k_B\equiv 1$) $T=0$, with environment squeezing parameter
$r=0$ and $1$, respectively.  The small-dashed and large-dashed curves
correspond to temperature $T=5$, with $r=0$ and 1, respectively.}
\label{fig:alfa}
\end{figure}

Figure \ref{fig:nu} is  a representative plot, showing that 
for large bath exposure time, $\nu(t)$ approaches 1. 
This Figure also brings out the
concurrent  behaviour of  temperature  and squeezing  with respect  to
$\nu(t)$. At large $t$,  $\alpha(t)$ also  approaches unity.   However,
unlike  the  case of  $\nu(t)$,  temperature  and  squeezing can  have  a
contrastive   effect   on   $\alpha(t)$,   as   brought   out   in   Fig.
\ref{fig:alfa}.  This
contrastive effect of squeezing with respect to time has been observed
in the  case of mixed  state geometric phase \cite{srigp}  and quantum
phase    diffusion    \cite{sriph}.    The   dot-dashed    curve    in
Fig.  \ref{fig:alfa}  represents  a  squeezed  amplitude  damping
channel,  i.e.,  a  channel  given  by {\em zero}  temperature  but  finite
squeezing.

\begin{figure}
\includegraphics[width=8.0cm]{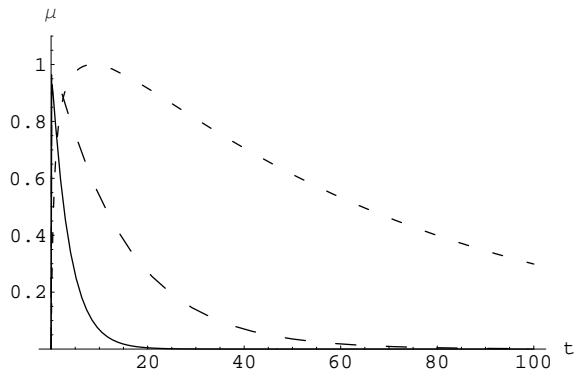}
\caption{$\mu(t)$ [Eq. (\ref{eq:mu})] as  a function of time $t$, with
$\gamma_0=0.05$  and  $r=1$.  The asymptotic value of $\mu(t)$ is 0.
The  solid, large-dashed and small-dashed curves
correspond to temperature (in units where $\hbar \equiv k_B \equiv 1$)
$T$ equals 20, 5 and 1, respectively.}
\label{fig:mu}
\end{figure}

The fact that  as time progresses, squeezing effects  tend to die out,
leaving thermal effects  alone to govern the system 
evolution, is illustrated in 
Fig.    \ref{fig:mu}.   Squeezing   of  the   bath   modes  introduces
non-stationary effects  due to  correlations between the  modes.  This
Figure also  shows the  washing out of  these effects with  time being
accentuated with increase in temperature.

\begin{figure}
\includegraphics[width=8.0cm]{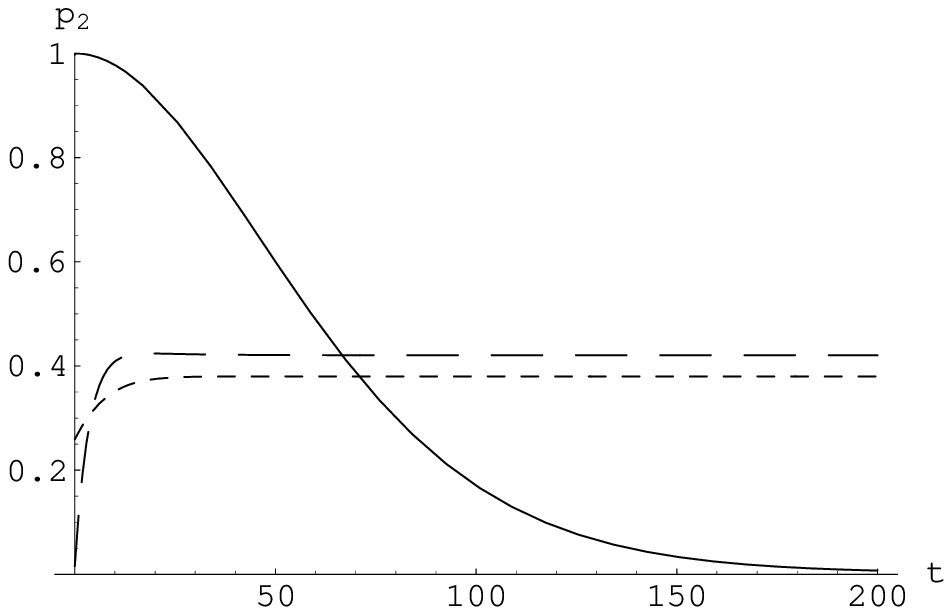}
\caption{Probability $p_2$  [Eq. (\ref{eq:p2})] as a  function of time
$t$   approaches   the    asymptotic   value   of   $N/(2N+1)$.   Here
$\gamma_0=0.05$.  The solid curve corresponds to temperature (in units
where  $\hbar   \equiv  k_B  \equiv   1$)  $T=0$  and   $r=0.05$.  The
small-dashed  and large-dashed  curves correspond  to $T=2$,  with $r$
equal  to 0.1  and 0.5,  respectively.  We  note that  the  solid line
depicts the  transformation of the squeezed  amplitude damping channel
to the amplitude damping channel.}
\label{fig:p2}
\end{figure}
Unlike the case of the generalized amplitude damping channel, here the
probabilities $p_1(t)$  and $p_2(t)$ are time-dependent  on account of
the  presence  of  non-stationary   effects  introduced  by  the  bath
squeezing  (Fig.  \ref{fig:p2}),  and  $p_2(t)$ eventually  reaches  a
stationary   value   of   $N/(2N+1)$,   as  may   be   inferred   from
Eqs. (\ref{eq:p2}) and (\ref{eq:auxip2}). Substituting this asymptotic
value  in Eqs.   (\ref{eq:nu}), (\ref{eq:mu})  and  (\ref{eq:tau}), we
find $\nu(\infty)=1$,  $\mu(\infty)=0$ and $\alpha(\infty)=1$,  as was
seen   in  Figs.   \ref{fig:nu},   \ref{fig:mu}  and   \ref{fig:alfa},
respectively.

In   the   absence  of   squeezing,   $p_2(\infty)$  becomes   $N_{\rm
th}/(2N_{\rm  th}+1)$,  consistent  with  the expression  for  $p$  in
Eq. (\ref{eq:lambda})  for the generalized  amplitude damping channel.
The  solid  line in  Fig.  \ref{fig:p2}  corresponds  to the  squeezed
amplitude damping channel,  which for the case of  zero bath squeezing
yields the action of a  quantum deleter \cite{qdele} via the amplitude
damping channel.

If   we   have   $a=0$ and $T=0$,   then  $\mu(t|a=T=0)=0$,   as   seen   from
Eq.  (\ref{eq:mu}), and  $p_2(t)=0$  because $p_2(t)$  in Eq.  (\ref{eq:p2})
reduces    to   $N_{\rm    th}/(2N_{\rm    th}+1)$   when    squeezing
vanishes.     Further,    $\nu(t|a=T=0)    =     \lambda(t|T=0)$    by
Eq. (\ref{eq:ha}).  Substituting these values  in Eqs. (\ref{eq:new0}),
we obtain  the amplitude  damping channel.  Eqs.  (\ref{eq:new0}) thus
furnish a complete representation  of a squeezed generalized amplitude
damping channel.

\section{Classical capacity of a squeezed generalized amplitude
damping  channel \label{sec:cc}}  

\begin{figure}
\includegraphics[width=8.0cm]{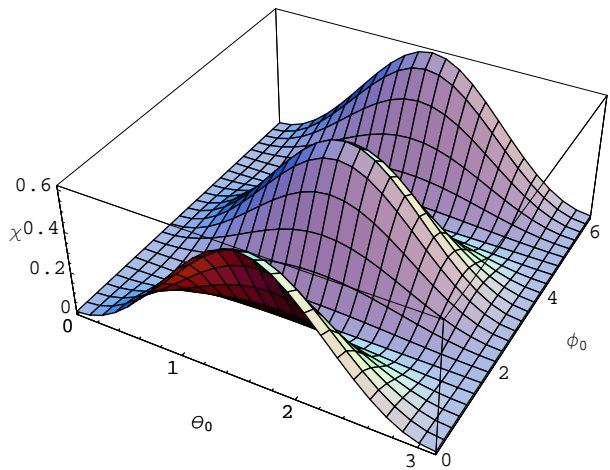}
\caption{Plotting the Holevo  bound $\chi$ [Eq. (\ref{eq:holevo})] for
a squeezed  amplitude damping channel with $\Phi=0$  and $f=0.5$, over
the  set  $\{\theta_0,\phi_0\}$, which  parametrizes  the ensemble  of
input   states  $\{(\theta_0,\phi_0),~(\theta_0+\pi,\phi_0)\}$.   Here
temperature  (in  units  where  $\hbar\equiv  k_B  \equiv  1$)  $T=5$,
$\gamma_0=0.05$, time $t=5.0$ and bath squeezing parameter $r=1$.  The
channel capacity  $C$ is  seen to correspond  to the optimal  value of
$\theta_0=\pi/2$ [i.e., the input states $\frac{1}{\sqrt{2}}(|0\rangle
\pm |1\rangle)$ for $\phi_0=0$].}
\label{fig:cc}
\end{figure}

A quantum  communication channel  can be used  to perform a  number of
tasks,  including transmitting  classical or  quantum  information, as
well as  for the cryptographic purpose of  creating shared information
between  a  sender  and  receiver,  that is  reliably  secret  from  a
malevolent eavesdropper \cite{sw98}.  A  natural question is as to how
information communicated over a squeezed generalized amplitude damping
channel (denoted ${\cal E})$, and given in the Kraus representation by
Eq.  (\ref{eq:new0}),  is  degraded.   In  this  Section,  we  briefly
consider the communication of classical information across the channel
\cite{sw97}.  The problem can be  stated as the following game between
sender  Alice and  receiver  Bob: Alice  has  a classical  information
source   producing  symbols  $X   =  0,\cdots,n$   with  probabilities
$p_0,\cdots,p_n$.  She encodes the  symbols as quantum states $\rho_j$
($0  \le  j  \le n$)  and  communicates  them  to Bob,  whose  optimal
measurement  strategy maximizes his  accessible information,  which is
bounded above by the Holevo quantity
\begin{equation}
\label{eq:holevo}
\chi = S(\rho) - \sum_j p_j S(\rho_j),
\end{equation}
where $\rho  = \sum_j  p_j \rho_j$, and  $\rho_j$ are  various initial
states \cite{holevo}.   
In the  present case,  we  assume Alice  encodes her  binary
symbols of  0 and 1  in terms of  pure, orthogonal states of  the form
(\ref{4b6}), and transmits them over the squeezed generalized
amplitude damping channel.

We further assume that Alice transmits her messages as product states,
i.e., without entangling them  across multiple channel use.  Then, the
(product  state) classical  capacity  $C$ of  the  quantum channel  is
defined  as  the  maximum  of  $\chi({\cal  E})$  over  all  ensembles
$\{p_j,\rho_j\}$ of possible input states $\rho_j$ \cite{hau96,sww96}.
In  Fig.   \ref{fig:cc},  we  plot  $\chi({\cal  E})$  over  pairs  of
orthogonal       input       states      $(\theta_0,\phi_0)$       and
$(\theta_0+\pi,\phi_0)\}$, which  correspond to  the symbols 0  and 1,
respectively, with  probability of the  input symbol 0  being $f=0.5$.
Here we take $\Phi=0$, and the optimum coding is seen to correspond to
the choice $(\theta_0=\pi/2,\phi_0=n\pi)$, where  $n \in I$, i.e., the
input   states   $\frac{1}{\sqrt{2}}(|0\rangle   \pm  |1\rangle)$   or
$\frac{1}{\sqrt{2}}(|0\rangle \mp |1\rangle)$.

\begin{figure}
\includegraphics[width=8.0cm]{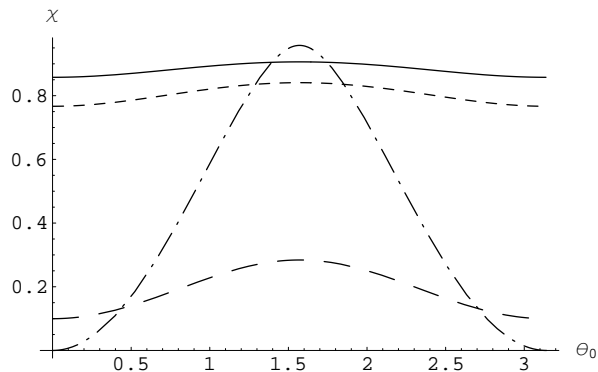}
\caption{Optimal  source  coding for  the  squeezed amplitude  damping
channel, with  $\chi$ plotted against $\theta_0$  corresponding to the
``0" symbol.  Here $\Phi=0$, $\gamma_0=0.05$ and $f=0.5$.   It is seen
that $\chi$ is  maximized for states of the  form (\ref{4b6}) when the
pair of input  states are given by $(\theta_0=\frac{\pi}{2},\phi_0=0)$
and    $(\theta_0=\frac{\pi}{2}+\pi,    \phi_0=0)$    [i.e.,    states
$\frac{1}{\sqrt{2}}(|0\rangle   \pm  |1\rangle)$].    The   solid  and
small-dashed curves represent temperature (in units where $\hbar\equiv
k_B \equiv 1$) $T=0$ and bath squeezing parameter $r=0$, but $t=1$ and
2,  respectively.  The large-dashed  and  dot-dashed curves  represent
$T=5$ and $t=2$,  but with $r=0$ and 2,  respectively. A comparison of
the  solid  and small-dashed  (small-dashed  and large-dashed)  curves
demonstrates   the  expected  degrading   effect  on   the  accessible
information,  of increasing  the  bath exposure  time $t$  (increasing
$T$).   A  comparison  of   the  large-dashed  and  dot-dashed  curves
demonstrates   the  dramatic  effect   of  including   squeezing.   In
particular, whereas squeezing  improves the accessible information for
the   pair   of   input   states   $\frac{1}{\sqrt{2}}(|0\rangle   \pm
|1\rangle)$, it  is detrimental for input  states $(\theta_0, \phi_0)$
given   by   $(0,0)$   (i.e.,   $|1\rangle$)  and   $(\pi,0)$   (i.e.,
$|0\rangle$).}
\label{fig:holevo_theta}
\end{figure}

\begin{figure}
\includegraphics[width=8.0cm]{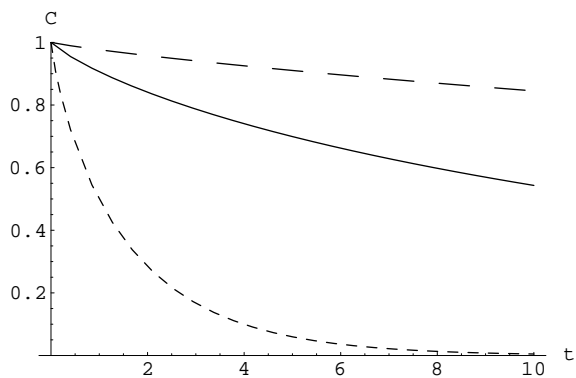}
\caption{Interplay  of  squeezing  and  temperature on  the  classical
capacity  $C$ of the  squeezed amplitude  damping channel  (with input
states  $\frac{1}{\sqrt{2}}(|0\rangle  \pm  |1\rangle)$, and  $f=1/2$,
corresponding   to   the   optimal   coding).    Here   $\Phi=0$   and
$\gamma_0=0.05$.  The solid and small-dashed curves correspond to zero
squeezing $r$, and temperature (in units where $\hbar\equiv k_B \equiv
1$) $T =  0$ and 5, respectively.  The  large-dashed curve corresponds
to $T=5$  and $r=2$. A  comparison between the solid  and large-dashed
curves shows that squeezing can improve $C$.}
\label{fig:holevo_sqtemp}
\end{figure}

Fig.   \ref{fig:holevo_theta}  depicts  $\chi({\cal E})$  for  various
channel parameters, with the pair  of orthogonal input states given by
$(\theta_0,\phi_0=0)$  and   $(\theta_0+\pi,\phi_0=0)$.  As  expected,
longer  exposure  to  the  channel, or  higher  temperature,  degrades
information more, but  the optimal choice of input  states remains the
same  as  before.  Interestingly,  squeezing improves  the  accessible
information for  input states  in a certain  range of  $\theta_0$, but
impairs it  in other. This  is consistent with the  understanding that
the           benefits            of           squeezing           are
quadrature-dependent.  Fig.  \ref{fig:holevo_sqtemp} demonstrates  the
contrastive effects  of temperature  and squeezing on  $C$.  Comparing
the solid and small-dashed curves, one notes that thermal effects tend
to  degrade $C$, whereas  bath squeezing  can improve  it, as  seen by
comparing  the   small-  and   large-dashed  curves.   In   fact,  the
improvement  due  to  squeezing  is  brought  out  dramatically  by  a
comparison of the solid  and large-dashed curves.  This highlights the
possible usefulness of squeezing to noisy quantum communication.

\section{Conclusions \label{sec:konklu}}

In this paper we have have  obtained a Kraus representation of a noisy
channel,  which we  call  the squeezed  generalized amplitude  damping
channel,  corresponding  to  the  interaction of  a  two-level  system
(qubit) with  a squeezed thermal  bath via a  dissipative interaction.
The  resulting  dynamics,   governed  by  a  Lindblad-type  evolution,
generates  a completely  positive map  that extends  the concept  of a
generalized  amplitude   damping  channel,  which   corresponds  to  a
dissipative  interaction  with a  purely  thermal  bath. The  physical
motivation for studying this channel  is that using a squeezed thermal
bath the decay rate of quantum coherence can be suppressed, leading to
preservation of nonclassical effects.  This is in contrast to the case
of a  purely dephasing  channel, where the  action of  squeezing, like
temperature, tends to  decohere the system.  We studied  the
characteristics of the squeezed generalized amplitude damping channel,
including its classical  capacity $C$.  We showed that  as a result of
bath  squeezing, it is  possible by  a judicious  choice of  the input
states, to  improve $C$ over the corresponding  unsqueezed case.  This
could have interesting implications for quantum communication.

\end{document}